\documentstyle[11pt]{article}

\newdimen\paperwidth \newdimen\paperlength \newdimen\margin
\newdimen\vmargin
\textwidth=\paperwidth \advance \textwidth by -2\margin
\advance\hoffset by -1truein \advance\hoffset by \margin
\textheight=\paperlength \advance \textheight by -2\vmargin
\advance\voffset by -1truein \advance\voffset by \vmargin
\advance\textheight by -2\baselineskip
\begin{document}

\renewcommand{\theequation}{\thesection.\arabic{equation}}
\newcommand{\Section}[1]{\section{#1}\setcounter{equation}{0}}

\begin{titlepage} \title{ {\bf   Analytic Formulations of the}\\
 {\bf Density Matrix Renormalization Group}\thanks{Work partly
supported by CICYT under contracts AEN93-0776 (M.A.M.-D.) and
PB92-109 ,
 European Community Grant ERBCHRXCT920069 (G.S.).} }

\vspace{2cm}    \author{ {\bf Miguel A. Mart\'{\i}n-Delgado}\dag
\mbox{$\:$} and {\bf Germ\'an Sierra}\ddag \\ \mbox{}    \\ \dag{\em
Departamento de F\'{\i}sica Te\'orica I}\\ {\em Universidad
Complutense.  28040-Madrid, Spain }\\ \ddag{\em Instituto de
Matem\'aticas y F\'{\i}sica Fundamental. C.S.I.C.}\\ {\em Serrano
123, 28006-Madrid, Spain } } \vspace{5cm}
 \date{} \maketitle \def\baselinestretch{1.3} \begin{abstract}

We present two new analytic formulations of the Density Matrix
Renormalization Group Method. In these formulations we combine the
block renormalization group (BRG) procedure with Variational and
Fokker-Planck methods. The BRG method is used to reduce the  lattice
size while the latter are used to construct approximate target states
to compute the block density matrix. We apply our DMRG methods to the
Ising Model in a transverse field  (ITF model) and compute several
of its critical properties which are then compared with the  old BRG
results. \ \

\ \

\ \

\ \

\ \

\end{abstract}

\vspace{2cm} PACS numbers: 75.10 Jm, 05.50.+q, 64.60.Ak

\vskip-17.0cm \rightline{UCM/CSIC-95-09} \rightline{{\bf September
1995}} \vskip3in \end{titlepage}

\newpage

\section{Introduction}

The Density Matrix Renormalization Group Method (DMRG) introduced by
White \cite{white}  has arguably become the most powerful numerical
tool to retrieve the essential features  of interacting quantum
Hamiltonians in 1D such as spin systems \cite{white},
\cite{white-huse} and fermion systems \cite{white-scalapino}. This
method  is a real-space renormalization group (RG) method which is
specially well suited when dealing  with zero temperature properties
of many-body systems, a situation where the Quantum  MonteCarlo
methods happen to be particularly badly behaved as far as fermionic
systems are concerned \cite{hirsch}.

 The origin of the density matrix RG method relies on the special
treatment carried out  by White and Noack \cite{white-noack} on the
1D tight-binding model,  the lattice version of a single particle in
a box. It was Wilson the first to point out the relevance  of this
simple model in understanding the sometimes bad numerical performance
of the  standard Block Renormalization Group (BRG) method. In
reference \cite{white-noack} the  authors proposed a method called
Combination of Boundary Conditions (CBC) which  performs extremely
well as compared to the exact known solution of the model.  Recently,
we have clarified the role played by the boundary conditions in the
real-space  renormalization group method \cite{bc-germanyo} by
constructing a new analytical BRG-method which is able to give the
exact ground state  of the model and the correct $1/N^2$-law for the
energy of the first excited state in the  large $N$(size)-limit.

\noindent The problem with the CBC method is the difficulty when
trying to  generalize it to  interacting models. In
\cite{white-noack} yet another method was presented called the
Superblock Method which is the precursor of the density matrix RG
method. The DMRG method is  able to treat many-body problems as for
example the 1D Heisenberg model of spin $S=1$  \cite{white},
\cite{white-huse} which happens to be a non-integrable model.

\noindent In this paper we present two novel versions of the DMRG
method based upon  the Perturbative-Variational and Fokker-Planck
approaches to quantum lattice Hamiltonians recently introduced in
references \cite{german-esteve} and \cite{german-fernando}
respectively. We arrived at these new  methods by searching for an
analytical formulation of the  density matrix RG method. As it
happens, the usual implementations of the DMRG method  are
intrinsically numerical  for they rely on Wilson's procedure of
enlarging the system size in his RG-treatment of the Kondo problem
\cite{wilson}. We find interesting to address the  problem of
constructing analytical extensions of the DMRG method for several
reasons. Firstly, it is known that standard block RG methods proved
to be useful when dealing with  qualitative features of some
important models such as the ITF model \cite{drell}, Lattice Gauge
Model \cite{fradkin},  Heisenberg model \cite{rabin}, Hubbard model
\cite{hirschII}, etc. and we want to see how the DMRG method performs
when compared to those analytical BRG treatments. To this purpose, we
have applied our Variational and Fokker-Planck DMRG methods to  the
Ising model in a tranverse field (ITF model). There are more
interesting models, but
 the ITF model is simple enough for a first application of these
methods. As a matter of fact, it was Drell and the SLAC group
\cite{drell} who started to apply the standard Block Renormalization
Group menthod to study QCD and they used the ITF  model as a test
model. Later on in the 80's this BRG method has been also applied to
the study of strongly correlated systems and its implications in
High-$T_c$ Superconductivity.

\noindent Secondly, the DMRG method as it stands is also
intrinsically one-dimensional and it is an  open problem to find
feasible numerical schemes to work with higher dimensional systems.
As it happens, our Variational and Fokker-Planck DMRG methods are
generalizable to  dimension higher than one and we might consider
them as a first attempt to solving this  important extension of the
DMRG method.

\noindent This paper is organized as follows. In Sect.2 we present a
brief introduction to the Block Renormalization Group methods based
upon the concept of  the intertwiner operator $T$. This allow us to
make a unified formulation of both the standard BRG methods and the
new Density Matrix RG method according to our analytic formulation.
In Sect.3 the Block RG-method is applied to the ITF model following
the formulation of Sect.2 and several critical exponents are computed.
In Sect.4 we present our new DMRG methods and apply them to  compute
the intertwiner operator and critical exponents for the  ITF model.
The results are compared with the old BRG results and  we find that
the density matrix methods perform better. Sect.5 is devoted to
conclusions and prospectives.

\vspace{30 pt}

\section{Block Renormalization Group Methods (BRG): Brief Review}

In this section the block renormalization group method is revisited
and we present a new  and unified reformulation of it based on the
idea of the {\em intertwiner operator} $T$
 to be discussed  below. This formulation will allow us to introduce
the new Variational and Fokker-Planck  DMRG methods on equal footing
as the standard BRG method. For a more extensive account on this
method we refer to \cite{jullienlibro}  and chapter 11 of reference
\cite{jaitisi} and references therein.

The block RG-method is a real-space RG-method introduced and
developed by the SLAC  group \cite{drell}. Let us recall that Wilson
developed his numerical  real-space renormalization group procedure
to solve the  Kondo problem \cite{wilson}.  It was clear from the
beginning that one could not hope to achieve the accuracy  Wilson
obtained for the Kondo problem when dealing with more complicated
many-body  quantum Hamiltonians as the ones mentioned in the
introduction. The {\em key difference} is  that in the Kondo model
there exists  a {\em recursion relation} for Hamiltonians at each
step  of the RG-elimination of degrees of freedom. Squematically,

\begin{equation} H_{N+1} = H_N + \mbox{hopping boundary term}
\label{d1} \end{equation}

\begin{equation} H_{N+1} = R(H_N)        \label{d1b} \end{equation}

\noindent The existence of such recursion relation facilitates
enormously the work, but as it  happens it is specific of {\em
impurity problems}.

{}From the numerical point of view, the Block Renormalization Group
procedure proved to be not  fully reliable in the past particularly
in comparison with other numerical approaches, such as  the Quantum
MonteCarlo method which were being developed at the same time. This
was one of  the reasons why the BRG methods remained undeveloped
during the '80's until the begining of  the '90's when they are
making a comeback as one of the most powerful numerical tools  when
dealing with zero temperature properties of many-body systems.

Let us first summarize the main features of the real-space RG. The
problem that one faces   generically is that of diagonalizing a
quantum lattice Hamiltonian $H$, i.e.,

\begin{equation} H |\psi > = E |\psi >
\label{1} \end{equation}

\noindent  where $|\psi >$ is a state in the Hilbert space $\cal H$.
If the lattice has N sites  and there are $k$ possible states per
site then the dimension of $\cal H$ is simply

\begin{equation} dim{ \cal H} = k^N
\label{2}
 \end{equation}

\noindent As a matter of illustration we cite the following
examples:  $k=4$  (Hubbard model), $k=3$ (t-J model), $k=2$
(Heisenberg model) etc.

\noindent When $N$ is large enough the eigenvalue problem (\ref{1})
is out of the capability of  any human or computer means unless the
model turns out to be integrable which only happens in some instances
in $d=1$.

\noindent  These facts open the door to a variety of approximate
methods among which the RG-approach,  specially when combined with
other techniques (e.g. numerical, variational etc.), is one of the
most relevant. The main idea of the RG-method is the mode elimination
or thinning of the degrees  of freedom followed by an iteration which
reduces the number of variables step by step until  a more manageable
situation is reached. These intuitive ideas give rise to a well
defined  mathematical description of the RG-approach to the low lying
spectrum of quantum lattice  hamiltonians.

To carry out the RG-program it will be useful to introduce the
following objects:

\begin{itemize}

\item ${\cal H}  $ : Hilbert space of the original problem.

\item ${\cal H }' $: Hilbert space of the effective degrees of
freedom.

\item $H $: Hamiltonian acting in $\cal H$.

\item $H'$: Hamiltonian acting in $\cal H'$ (effective Hamiltonian).

\item $T  $ : embedding operator : ${\cal H }'\longrightarrow {\cal
H}$

\item $T^{\dagger }  $ :truncation operator : ${\cal H}
\longrightarrow {\cal H}'$

\end{itemize}

The problem now is to relate $H$, $H'$ and $T$. The criterium to
accomplish this task is that  $H$ and $H'$ have in common their low
lying spectrum. An exact implementation of this  is given by the
following equation:

\begin{equation} H T = T
H'                                                   \label{3}
\end{equation}

\noindent which imply that if $\Psi '_{E'}$ is an eigenstate of $H'$
then  $T\Psi '_{E'}$ is an eigenstate of $H$ with the same eigenvalue
(unless it belongs to the kernel of $T$: $T\Psi '_{E'}=0$), indeed,

\begin{equation} HT\Psi '_{E'} =  T H'\Psi '_{E'} = E' T\Psi
'_{E'}                                                  \label{4}
\end{equation}

To avoid the possibility that $T\Psi '=0$ with $\Psi '\neq 0$, we
shall impose on $T$ the condition,

\begin{equation} T^{\dagger } T = 1_{\cal
H'}                                                   \label{5}
\end{equation}

\noindent such that

\begin{equation} \Psi = T\Psi '  \Rightarrow \Psi ' = T^{\dag}
\Psi                                                  \label{6}
\end{equation}

\noindent Condition (\ref{5}) thus stablishes a one to one relation
between $\cal H'$ and Im($T$)  in $\cal H$.

\noindent Observe that Eq. (\ref{3}) is nothing but the commutativity
of the following  diagram:

\begin{center} \[ \begin{array}{llcll}
  & {\cal H'} & \stackrel{T}{\longrightarrow} & {\cal H}  &  \\
 H' & \downarrow &     &  \downarrow & H  \\
  &  {\cal H'}  &  \stackrel{T}{\longrightarrow} & {\cal H} &
\end{array} \] \end{center}

 Eqs. (\ref{3}) and (\ref{5}) characterize what may be called exact
renormalization group method  (ERG) in the sense that  the whole
spectrum of $H'$ is mapped onto a part (usually the bottom part)  of
the spectrum of $H$. In practical cases though the exact solution of
Eqs. (\ref{3}) and (\ref{5})  is not possible so that one has to
resort to approximations (see later on). Considering  Eqs. (\ref{3})
and (\ref{5}) we can set up the effective Hamiltonian $H'$ as:

 \begin{equation} H' = T^{\dag} H
T                                                   \label{10}
\end{equation}

\noindent This equation does not imply that the eigenvectors of  $H'$
are mapped onto eigenvectors  of $H$. Notice that Eq.(\ref{10})
together with (\ref{5}) does not imply Eq. (\ref{3}). This happens
because the converse of Eq.(\ref{5}), namely $TT^{\dag } \neq 1_{\cal
H}$  is not true, since otherwise this equation together with
(\ref{5}) would imply that the Hilbert spaces {\cal H} and {\cal H'}
are isomorphic while on the other  hand the truncation inherent to
the RG method assumes that dim {\cal H'} < dim {\cal H}.

What Eq.(\ref{10}) really implies is that the mean energy of $H'$ for
the states $\Psi '$ of $\cal H'$  coincides with the mean energy of
$H$ for those states of $\cal H$ obtained through the embedding  $T$,
namely,

  \begin{equation} <\Psi '|H'|\Psi'> =  <T\Psi ' |H| T\Psi'>
\label{11} \end{equation}

In other words $ T\Psi'$ is used as a variational state for the
eigenstates of the Hamiltonian $H$. In  particular $T$ should be
chosen in such a way  that the states truncated in $\cal H$ , which
go down  to $\cal H'$, are the ones expected to contribute the most
to the ground state of $H$. Thus Eq.  (\ref{10}) is the basis of the
so called variational renormalization group method (VRG)
\footnote{The word variational here is used with a different meaning
as in the  introduction of this paper where it refers to the
variational choice of the target state in the  DMRG method to be
discussed in section 4.}. As a matter  of fact, the VRG method was
the first one to be proposed.  The ERG came afterwards as a
perturbative extension of the former (see later on).

\noindent  More generally, any operator $\cal O$ acting in $\cal H$
can be ``pushed down" or renormalized to a new  operator $\cal O'$
which acts in $\cal H'$ defined by the formula,

  \begin{equation} {\cal O}'  = T^{\dag} {\cal O}
T                                       \label{43} \end{equation}

\noindent Notice that Eq.(\ref{10}) is a particular case of this
equation if choose {\cal O} to  be the Hamiltonian $H$.

In so far we have not made use of the all important concept of the
block,  but a practical implementation of the VRG or ERG methods does
require it. The central role played by this concept  makes all the
real-space RG-methods to be block methods.

Once we have established the main features of the RG-program,  there
is quite freedom to implement specifically these fundamentals. We may
classify this freedom in two aspects:

\begin{itemize}

\item The choice of how to reduce the size of the lattice.

\item The choice of how many states to be retained in the truncation
procedure.

\end{itemize}

\noindent We shall address the first aspect now. There are mainly two
procedures to reduce the  size of the lattice:

\begin{itemize}

\item by dividing the lattice into blocks with $n_s$ sites each. This
is the blocking method  introduced by Kadanoff to treat spin lattice
systems.

\item by retrieving site by site of the lattice at each step of the
RG-program. This is the procedure used by Wilson in his RG-treatment
of the Kondo problem. This method is clearly more suitable  when the
lattice is one-dimensional.

\end{itemize}

  We shall be dealing with the Kadanoff
 block methods mainly because they are well suited to perform
analytical computations and because they are conceptually easy to be
extended to higher  dimensions. On the contrary, the DMRG method
introduced by White \cite{white}  works with the Wilsonian numerical
RG-procedure what makes it intrinsically one-dimensional  and
difficult to be generalized to more dimensions. Thus why we shall
formulate our  Variational and Fokker-Planck DMRG procedures as block
renormalization methods  in section 4.

\noindent Block RG-methods have recently received also renewed
attention in one-dimensional problems in connection to what is called
a {\em quantum group} symmetry \cite{q-germanyo}. Based upon this
symmetry we have constructed a new BRG-method that we call $q$-RG
which among other features it is able to predict the exact line of
critical XXZ models in the  Anisotropic Heisenberg model, unlike the
standard BRG-method.

To exemplify the standard BRG-method we shall study a 1d-lattice
Hamiltonian,  the Ising model in a transverse field (ITF model). The
main ideas are also valid in  higher dimensions although computations
are more involved.  Hence we shall be dealing with a one-dimensional
lattice, usually a periodic chain.
 In every site of the chain there are $k$ degrees of freedom, hence:

 \begin{equation} {\cal H} = {\cal C} ^k \otimes
\stackrel{N}{\ldots}\otimes {\cal C} ^k := \otimes ^N {\cal C} ^k
\label{12} \end{equation}

\noindent We shall consider Hamiltonians $H$ containing operators
which involve only a single-site  part $H_{\mbox{s}}$ or
two-nearest-neighbour-site part $H_{\mbox{ss}}$ and will be
simbolically  depicted as in Fig.1, in such a way that,

  \begin{equation}
 H =    H_{\mbox{s}} + H_{\mbox{ss}}                  \label{14}
\end{equation}

\noindent As a matter of illustration, let us give one example of
this decomposition  in the ITF model which will turn out  to be very
useful in putting many key ideas to the test.

\section{Block RG-Approach to the Ising Model in a Transverse Field
(ITF)}

The Ising Model in a Transverse Field is originally a one-dimensional
quantum lattice system  with quantum critical properties equal to the
well-known thermal critical properties of the  classical 2D-Ising
Model. The lattice Hamiltonian of the ITF model is:

   \begin{equation} H_N (\Gamma ,J) = -\Gamma \sum _{j=1}^N \sigma
_j^x
 - J \sum _{j=1}^N \sigma _j^z \sigma
_{j+1}^z                                     \label{46} \end{equation}

\noindent where $\sigma _j^x$ and $\sigma _j^z$ are the standard
Pauli matrices acting at the  $j$-th site of the chain.

The Hilbert space of states and the intrablock and interblock
Hamiltonians for this model are, respectively:

  \begin{equation} {\cal H} = \otimes_1^N {\cal C}
^2                  \label{15a} \end{equation}

  \begin{equation}
   H_{\mbox{s}} = -\Gamma \sum _{j=1}^N \sigma _j^x
\label{15b} \end{equation}

  \begin{equation}
 H_{\mbox{ss}} = -J \sum _{j=1}^N \sigma _j^z \sigma
_{j+1}^z                    \label{15c} \end{equation}

The first step of the BRG method consists in asembling the set of
lattice points into diconnected blocks of $n_B$ sites each, as in
Fig.2.

In this fashion there are a total of $N'=N/n_B$ blocks in the whole
chain. This partition of the lattice into blocks induces a
decomposition of the Hamiltonian (\ref{14})  into an intrablock
Hamiltonian  $H_B$ and a interblock Hamiltonian $H_{BB}$ as
illustrated in Fig.3.

\noindent Observe that the block Hamiltonian $H_B$  is a sum of
commuting Hamiltonians each  acting on every block. The
diagonalization of $H_B$ can thus be achieved for small $n_B$ either
analytically or numerically. The content of Fig. 3 can be written as

 \begin{equation}
 H = H_B + \lambda H_{BB}                    \label{21} \end{equation}

\noindent where $\lambda $ is a coupling constant which is already
present in $H$ or else it can  be introduced as a parameter
characterizing the interblock coupling and in this latter case one
can  set it to one at the end of the discussion.

\noindent Eq. (\ref{21}) suggests that we should search for solutions
of the intertwiner equation  (\ref{3}) in the form of a perturbative
expansion in the interblock coupling constant parameter  $\lambda $,
namely,

 \begin{equation} T = T_0 + \lambda T_1 +  \lambda^2  T_2 +
\ldots                    \label{22a} \end{equation}

 \begin{equation} H' =H'_0 + \lambda H'_1 +  \lambda^2 H'_2 +
\ldots                    \label{22b} \end{equation}

\noindent To zeroth order in $\lambda $ Eq. (\ref{3}) becomes

 \begin{equation} H_B T_0 = T_0
H_0'                                                 \label{23}
\end{equation}

\noindent Since $H_B$ is a sum of disconnected block Hamiltonians
$h_{j'}^{(B)}$, $j'=1,\ldots ,N'$ implicitly defined through the
relation

 \begin{equation} H_B = \sum
_{j'=1}^{N'}h_{j'}^{(B)}
\label{24} \end{equation}

\noindent one can search for a solution of $T_0$ in a factorized form

 \begin{equation} T_0 = \prod _{j'=1}^{N'}
T_{0,j'}                                                \label{25}
\end{equation}

\noindent and an effective Hamiltonian $H_0'$ which acts only at the
site $j'$ of the new chain,

 \begin{equation} H_0' =   \sum _{j'=1}^{N'}h_{j'}^{(s')} =
H'_{s'}                                                \label{26}
\end{equation}

\noindent Observe that $ H'_{s'}$ is nothing but a site-Hamiltonian
for the new chain. Eq. (\ref{23}) becomes for each block:

  \begin{equation} h_{j'}^{(B)}  T_{0,j'}   =
T_{0,j'}h_{j'}^{(s')}
\label{27} \end{equation}

\noindent The diagonalization of $h_{j'}^{(B)}  $ for $j'=1,\ldots
,N'$ will allow us to write

  \begin{equation} h_{j'}^{(B)} =  \sum _{i=1}^{k'} |i\rangle_{j'} \
\epsilon _i \ _{j'} \hspace*{-1pt} \langle i| +  \sum _{\alpha
=1}^{k^{n_s}-k'} |\alpha \rangle_{j'}  \ \epsilon _{\alpha }  \ _{j'}
\hspace*{-1pt} \langle \alpha |  \label{28} \end{equation}

\noindent where $|i>_{j'}$ for $j=1,\ldots ,k'$ are the $k'$-lowest
energy states of $h_{j'}^{(B)}$. Moreover, we suppose that
$h_{j'}^{(B)}$ is the same Hamiltonian for each block so that
$\epsilon _i$ does not depend on the block.

\noindent The truncated Hamiltonian $h_{j'}^{(s)}$ and the
intertwiner operator $T_{0,j'}$ are then  given by:

  \begin{equation} h_{j'}^{(s')} =  \sum _{i=1}^{k'} |i\rangle'_{j'}
\ \epsilon _i \ _{j'} \hspace*{-1pt} \langle i|'   \label{29}
\end{equation}

  \begin{equation} T_{0,j'} =  \sum _{i=1}^{k'} |i\rangle'_{j'} \
_{j'} \hspace*{-1pt} \langle i|'       \label{30} \end{equation}

\noindent  Later on we shall show examples of these relations.

To obtain the first order correction to the Hamiltonian $H'_1$ we
must consider Eq. (\ref{3}) to  first order in $\lambda$:

    \begin{equation} H_{BB} T_0 + H_B T_1 = T_0 H'_1 + T_1
H'_0                          \label{31} \end{equation}

\noindent Multiplying the left hand side by $T^{\dag }_0$ and using
$T^{\dag }_0T_0=1$ along with  $H_B T_0 = T_0 H'_0$ we readly obtain:

    \begin{equation} T^{\dag}_0H_{BB} T_0 + H'_0 T^{\dag}_0 T_1 =
H'_1 + T^{\dag}_0T_1 H'_0      \label{32} \end{equation}

We would like to kill the term proportional to $ T^{\dag}_0 T_1$. For
this purpose Eq. (\ref{5}) which  implies $ T^{\dag}_0 T_1+ T_1
T^{\dag}_0=1$ is not very useful. A resolution of this problem  can
be accomplished if instead of the operator $T$ one uses another
operator $\tilde{T}$ satisfying  the defining equations:

    \begin{equation} H \tilde{T} =    \tilde{T}
H'
\label{33a} \end{equation}

    \begin{equation} T^{\dag }_0 \tilde{T} = 1_{{\cal
H}'}                                         \label{33b}
\end{equation}

\noindent Then $ \tilde{T}_0=T_0$ and $ T^{\dag}_0 \tilde{T}_1=0$ in
which case Eq. (\ref{32}) simply becomes:

     \begin{equation} H'_1 = T^{\dag}_0H_{BB}T_0 =
H'_{s's'}                                         \label{34}
\end{equation}

We can summarize these results saying that up to first order in
$\lambda $, the effective Hamiltonian  $H'$ can be obtained using
simply the zeroth order intertwiner operator $T_0$ (see Fig.4):

     \begin{equation} H'_{\mbox{(up to order}\ \lambda )} = H'_{s'}
+  H'_{s's'}  = T^{\dag }_0  (H_B + \lambda H_{BB})
T_0
\label{35} \end{equation}

\noindent This is precisely the prescription of Drell et al.
\cite{drell}.

The second order correction to $H'$ can be obtained again from Eqs.
(\ref{33a})-(\ref{33b})
 and is given by

     \begin{equation} H'_{2 } = T^{\dag }_0 H_{BB}
\tilde{T}_1            \label{35b} \end{equation}

 There is a close parallelism between the perturbative solution of
Eqs. (\ref{3}) or (\ref{33a})-
 (\ref{33b}) and  the pertubation theory of the Schrodinger equation
for a Hamiltonian of the form $H_0 +  \lambda H_1$. As a matter of
fact, the normalization condition (\ref{33b}) for operators is
equivalent to the standard normalization for wavefunctions $<\Psi
_0|\Psi (\lambda )> =  <\Psi _0|\Psi _0> = 1$ that is adopted to
avoid normalization complications. In what follows  we shall mainly
concentrate on the first order solution Eq. (\ref{35}).

 The final outcome of this analysis is that the effective
Hamiltonian  $H'$ has a similar  structure to the one we started
with, $H$. The operators involved in $H'_{s'}$ and  $H'_{s's'}$ may
by all means differ from those of $H_{s}$ and $H_{ss}$, but in some
cases  the only difference shows up as a change in the coupling
constants. This is known as the  renormalization of the bare coupling
constants. When this is the case, one may easily iterate  the
RG-transformation and study the RG-flows.

Let us summarize the RG-prescription we have introduced so far in
Table 1.

\noindent We have denoted this prescription by BRG1($n_s$,$k'$) where
$n_s$ and $k'$  have been defined earlier and 1 denotes  that we are
working to first order in perturbation  theory.

For the ITF model we shall consider blocks of two sites ($n_s=2$)
and truncation to two states ($k'=2$). The block Hamiltonian has two
sites and has the form,

\begin{equation} h^{(B)} = -\Gamma (\sigma _1^x + \sigma _2^x ) - J
\sigma _1^z \sigma _2^z
  \ \ \ \ \ \    (\Gamma ,J > 0)\label{37} \end{equation}

\noindent The eigenstates of this block Hamiltonian (\ref{37}) are
given in increasing order of  energies by,

 \begin{equation} |G> = \frac{1}{\sqrt{1+a^2}} (|00> + a |11>)     \
\ \ \ E = -\sqrt{J^2 + 4 \Gamma ^2} \label{38a} \end{equation}

 \begin{equation} |E> = \frac{1}{\sqrt{2}} (|01> +  |10>)     \ \ \ \
\ \ \ E = -J              \label{38b} \end{equation}

 \begin{equation} |E'> = \frac{1}{\sqrt{2}} (|00> -  |11>)     \ \ \
\ \ \ \ E = J                     \label{38c} \end{equation}

 \begin{equation} |E''> = \frac{1}{\sqrt{1+a^2}} (-a |01>  +
|10>)     \ \ \ \ E = \sqrt{J^2 + 4 \Gamma ^2} \label{38d}
\end{equation}

\noindent $|0>$ and $|1>$ are the eigenstates of $\sigma ^x$,

 \begin{equation} \sigma ^x |0> = |0>, \ \ \ \ \sigma ^x |1> =
-|1>                              \label{39} \end{equation}

\noindent and $a = a(g)$ is the following function of the ratio
$J/2\Gamma := g$,

 \begin{equation} a(g) = \frac{-1 + \sqrt{1 +
g^2}}{g}                             \label{40} \end{equation}

\noindent which in turn satisfies

 \begin{equation} a(0) = 0, \ \ \ \  a(\infty) =
1                            \label{41} \end{equation}

\noindent The intertwiner operator whithin each block has the form

 \begin{equation} T_0(a) = |G\rangle \  \hspace*{-1pt} \langle 0|' +
|E\rangle \ \ _{} \hspace*{-1pt} \langle
1|'                                          \label{42} \end{equation}

\noindent where $|0>'$ and $|1>'$ form a basis of states at each
point of the new chain. The  effective Hamiltonian $H'$ up to order
$J$ can be computed from Eq. (\ref{35}).

\noindent Thus to get $\cal H'$ we have to study the renormalization
of the various operators  entering in its definition. Using
(\ref{42})  and (\ref{38a})-(\ref{38d})
 one obtains after some elementary  algebra:

   \begin{equation} T^{\dag} \sigma ^x_j T = \frac{1 - a^2}{2 (1 +
a^2)} (1 + \sigma ^x_{j'})   \label{44a} \end{equation}

   \begin{equation} T^{\dag} \sigma ^z_j T = \frac{1 + a}{\sqrt{2 (1
+ a^2)}} \sigma ^z_{j'}   \label{44b} \end{equation}

   \begin{equation} T^{\dag} \sigma ^z_{2j-1} \sigma ^z_{2j}  T =
\frac{(1 + a)^2}{2 (1 + a^2)} 1  +  \frac{(1 - a)^2}{2 (1 + a^2)}
\sigma ^x_{j'}                                    \label{44c}
\end{equation}

   \begin{equation} T^{\dag} \sigma ^z_{2j} \sigma ^z_{2j+1}  T =
\frac{(1 + a)^2}{2 (1 + a^2)} \sigma ^z_{j'} \sigma ^z_{j'+1}
\label{44d} \end{equation}

\noindent The range of the indexes run as follows:

   \begin{equation} j = 2j' - 1 + p     \label{45a} \end{equation}

   \begin{equation} j = 1,\ldots ,N  \label{45b} \end{equation}

   \begin{equation} j' = 1,\ldots ,N /2    \ \  \ \  \ \ \  p = 0,1
\label{45c} \end{equation}

\noindent Applying Eqs. (\ref{44a})-(\ref{44d}) to the ITF
Hamiltonian,  one gets

   \begin{equation} T^{\dag} H_N (\Gamma ,J) T =  \Delta E +
H_{N/2} (\Gamma ',J')   \label{47} \end{equation}

\noindent where

   \begin{equation} \Delta E = -\frac{N}{2} \left[ \Gamma \frac{1 -
a^2}{(1 + a^2)} +  \frac{J}{2}  \frac{(1 + a)^2}{(1 + a^2)}
\right]                                         \label{48a}
\end{equation}

   \begin{equation} \Gamma ' =  \Gamma \frac{1 - a^2}{(1 + a^2)} - J
\frac{(1 + a)^2}{2 (1 +
a^2)}                                           \label{48b}
\end{equation}

   \begin{equation} J' =  J \frac{(1 + a)^2}{2 (1 +
a^2)}                                           \label{48c}
\end{equation}

The derivation of Eqs. (\ref{44a})- (\ref{44d})
 and (\ref{47})-(\ref{48c})  does not make use of Eq. (\ref{40}) and
hence have a more general validity. In other words, we can use the
function $a(g)$ as a variational  function in order to construct
better ground states in the spirit of the VRG.

 \noindent Eq. (\ref{40}) is one of the numerous choices we can make.
We shall consider later on  other examples. Schematically we can set
up the following relationship,

   \begin{equation} \begin{array}{c} \mbox{RG-Prescription with }
n_s=2, k'=2 \end{array}    \Longleftrightarrow \left\{
\begin{array}{l}      a(g) \geq 0  \\ a(0) = 0, \ \ a(\infty) = 1
\end{array}  \right.
\label{49} \end{equation}

The physical properties of $H_N(\Gamma ,J)$ depend only upon the
ratio $g = J/2 \Gamma $. If $0 \leq g \leq 1/2$ one is in a
disordered region characterized by a unique ground state with
unbroken symmetry ($<\sigma ^z> = 0$). If $g > 1/2$ the ${\em Z} _2$
symmetry associated to the operator  $Q = \prod _j \sigma ^z_j$,
which commutes with $H_N $ for $N$ even, is broken. This is the
ordered phase which has two degenerate ground states corresponding
to  $<\sigma ^z> = \pm m \neq 0$.

 At $g = g_{\mbox{c}} = 1/2$ the system is critical and belongs to
the same universality class as the 2D-classical Ising model. The
critical exponents can be defined in terms of the behaviour of the
``quantum observables" as functions of $g_{\mbox{c}} - g$.

Most of the critical exponents can be  computed from the properties
of the RG-transformation. In the case of the ITF model the
 RG-transformation can be obtained from Eq.(\ref{48a})-(\ref{48c}):

   \begin{equation} g' := \frac{J'}{2 \Gamma '} := R(g) =
\frac{1}{2} \frac{g (1 + a(g))^2}{1 - a(g)^2 - g (1 - a(g))^2}
\label{50} \end{equation}

\noindent For any function $a(g)$ satisfying Eq. (\ref{49}), this
transformation has 3 fixed points  $g_{\ast } =  0, g_{\mbox{c}},
\infty$. The fixed points $g_{\ast} = 0$ and $\infty$ are attractive
and  correspond to the disordered and ordered phases respectively.
The fixed point at $g_{\mbox{c}}$ is repulsive and correspond to the
critical point of the ITF Hamiltonian. The value of the function
$a(g)$ at $g=g_{\mbox{c}}$ is a function only of $g_{\mbox{c}}$ and
does not depend on the  particular prescription chosen, that is, it
is a {\em universal function} :

   \begin{equation} g_{\mbox{c}} = R(g_{\mbox{c}}) \Rightarrow
a(g_{\mbox{c}}) = \frac{2 \sqrt{1 - 2 g_{\mbox{c}}} + 2 g_{\mbox{c}}
- 1} {3 + 2
g_c}
\label{51} \end{equation}

\noindent This equation implies in particular that whatever
prescription is chosen,  assuming that $a$ is real, the critical value
obtained from Eqs.(\ref{48a})-(\ref{48c})  will always be less than
the exact value $1/2$.

   \begin{equation} g_{\mbox{c}} \leq g_{\mbox{c}}^{\mbox{exact}} =
1/2   \label{52} \end{equation}

\noindent To get the value of $g_{\mbox{c}} $ for a given
prescription one has simply to find the  intersection of the function
$a(g)$ and the function,

   \begin{equation} f(g) = \left\{ \begin{array}{lll} \frac{2 \sqrt{1
- 2 g} + 2 g - 1}{3 + 2 g}  & &  0 \leq g \leq 1/2 \\ 0
&                   &                               g \geq 1/2
 \end{array}
\right.
\label{53} \end{equation}

\noindent as is shown in Fig.7.

The analysis of the RG-equations usually has to be done numerically.
However, there is a great  deal of information that can be retrieved
without completely solving the RG-equations if we  know wheather the
successive coupling constants $g_n \longrightarrow g_{n+1}=R(g_n)$
increase or decrease during the iteration procedure. To this end it
is convenient to introduce the the familiar beta function $\beta(g)$
of quantum field theory which in this context is \cite{drell},

   \begin{equation} \beta(g) := R(g) -
g                                                  \label{53b}
\end{equation}

\noindent In Fig.8  we have plotted the beta function for the ITF
model we are analyzing. A fixed  point of the transformation occurs
at values of $g$ which reproduce themselves under the  RG-iteration,
i.e., they are the zeroes of the beta function:

   \begin{equation} \beta( g_{\mbox{c}}) =
0                                                   \label{53c}
\end{equation}

\noindent There are 3 fixed points for besides the two zeroes  at
$g=0$ and $0.39$,  $g=\infty $ is also a fixed point for it cannot be
reduced by further iterations.

There is additional qualitative information which can be extracted
from the shape of the beta  function $\beta(g)$. In particular, the
sign of   $\beta(g)$ is responsible for the stability character of
the fixed point. When  $\beta(g)<0$ ($>0$) this means that $g$
decreases (increases) after  one iteration and the resulting $g'$
lies to the left (right) of the $g$ we started with.

  The outcome of this RG-analysis can be summarized by saying that
the fixed points  $g_{\ast}=0,\infty $ are {\em stable fixed points}
while $g_{c}=0.39$  is an {\em unstable fixed point}.

Given the RG-transformation Eq. (\ref{50}) we can compute several
critical exponents and compare  them with the exact results in order
to check the accuracy of the method.

\paragraph{Correlation Length Exponent $\nu$.}

It gives the behaviour of the correlation length in the vicinity of
$g_{c}$

   \begin{equation} \xi \sim (g -
g_{c})^{-\nu}
\label{54} \end{equation}

\noindent Under the RG-transformation (\ref{50}) $\xi \rightarrow
\xi' =\xi/2$ which leads inmediately  to an expression for $\nu$ in
terms of the derivative of $R(g)$ evaluated at the critical point,

    \begin{equation} \frac{1}{\nu } =  \frac{\ln R'(g_{c})}{\ln
2}                    \label{55} \end{equation}

\noindent From Eq. (\ref{50}) we can evaluate $R'(g_{c})$ as a
function of  $a_{c} = a(g_{c})$ and $a_{c} ' = \frac{da}{dg}|_{c}$ ,

    \begin{equation} \lambda_T :=  R'(g_{c}) = 1 + 2 g_{c}
 (\frac{1 - a_{c}}{1 + a_{c}})^2 +  4 \frac{a_{c}' g_{c} \sqrt{1 - 2
g_{c}}} {(1 +
a_{c})^2}
\label{56} \end{equation}

\noindent In Table  2 we show the value of $\nu $ obtained for
different choices of the function $a(g)$.

 \paragraph{Dynamical Exponent z.}

At the critical point where $g'=g$ holds, the Hamiltonian changes by
an overall factor which in turn  defines the dynamical exponent z

   \begin{equation} H_{c} \longrightarrow H_{c}' = \frac{1}{2^z}
H_{c}
\label{57} \end{equation}

\noindent In order to get z we notice that

  \begin{equation}
 \frac{1}{2^z} =  (\frac{J'}{J})_{c} =
 (\frac{\Gamma '}{\Gamma })_{c}
\label{58} \end{equation}

\noindent Hence

   \begin{equation} z  =  \frac{\ln (\frac{J'}{J})}{\ln 2} =  1 +
\frac{\ln \left[ (1 + a^2_{c})/(1 + a_{c})^2 \right]} {\ln
2}
\label{59} \end{equation}

It follows from Eq. (\ref{59}) and the positivity property of $
a_{c}$ (\ref{49}) that $z$ is always less than the exact value,

   \begin{equation} z  \leq z^{exact} =
1
\label{60} \end{equation}

 \paragraph{Magnetic Exponent $\beta $.}

This critical exponent is defined through the spontaneous
magnetization $M_z$ in the ordered phase,

   \begin{equation} M = <\sigma ^z_j>                      \label{61}
\end{equation}

\noindent Above $a_{\mbox{c}}$ but close to the critical point we
will have

   \begin{equation} M \sim (g -  g_{\mbox{c}})^{\beta
}                \label{62} \end{equation}

\paragraph{Gap Exponent $s$.} Using scaling arguments satisfied by
the BRG method, the gap $G$ behaves like

\begin{equation} G \sim (g- g_c) ^s \end{equation}

\noindent with

\begin{equation} s = \nu z \end{equation}

\noindent Equation (\ref{44b}) relates the magnetization $M$ and the
one obtained after the  RG-transformation

    \begin{equation} M= \frac{1 + a}{\sqrt{2 (1 + a^2)}}
M'              \label{63} \end{equation}

\noindent Combining Eqs.(\ref{63}), (\ref{62}) and $g'=R(g)$ we
arrive at,

    \begin{equation} \frac{M'}{M} = \left[   R'(g_{\mbox{c}})
\right]^{\beta } =  2^{\beta / \nu } = \frac{1 +
a_{\mbox{c}}}{\sqrt{2 (1 + a_{\mbox{c}}^2)}}            \label{64}
\end{equation}

\noindent Using Eq.(\ref{59}) we are able to relate the critical
exponents $\beta $, $\nu $ and $z$  through the following scaling
relation,

    \begin{equation} \beta = \frac{1}{2} z
\nu
\label{65} \end{equation}

\noindent This relation is valid for any choice of the function
$a(g)$ and therefore it is characteristic  of using a block
containing two sites. Observe that the exact exponents of the ITF
model never satisfy this  scaling relation (\ref{65}). This shows the
limitation of the block method when using a two-site block.

\section{Density Matrix RG Method: Analytic Formulation}

The Density Matrix RG-method (DMRG) is an improved version of the
real-space renormalization  group methods introduced by White
\cite{white} as a further elaboration of the ideas concerning the
Combination of Boundary Conditions method \cite{white-noack}.

The fundamental difficulty of the BRG method lies in choosing the
eigenstates of the block  Hamiltonian $H_B$ to be the states kept.
Since $H_B$ contains no connections to the rest  of the lattice its
eigenstates have inappropiate features at the block ends. The CBC
method of  White and Noack is a first attempt to solving this
intrinsic problem. The rationale of this  method was that quantum
fluctuations in the rest of the system effectively apply a variety
of  boundary conditions to the block. The CBC method proved to be
very effective for the simple  single-particle problem studied by
White and Noak, but it happens to be ill-suited to  interacting
systems. The importance of the CBC method relies more on the lessons
we can learn from it rather than  the specific technicalities
pertaining the simple case where it is applied successfully. There
are  two main ideas in order to proceed towards density matrix
RG-method, namely:

\begin{itemize}

\item The block is not isolated.

\item The eigenstates retained are not eigenstates of a unique block
Hamiltonian $H_B$.

\end{itemize}

 \noindent  The DMRG method is in a sense an evolution of the CBC
method in which we  ``let the system" to choose the best boundary
conditions. White suggests that for a system which is  {\em strongly
coupled} to the outside ``universe" (the rest of the lattice), it is
much more  appropiate to use {\em eigenstates of the block density
matrix} to describe the system  (block), rather than the eigenstates
of the system's Hamiltonian $H_B$.  White's proposal can be stated by
saying:

\begin{itemize}

\item Choose to keep the $n_B$ most probable eigenstates of the block
density matrix.

\end{itemize}

\noindent It is possible to show that keeping the most probable
eigenstates of the density matrix  gives the most accurate
representation of the state of the system as a whole \cite{white}.
This is the basis of the Density Matrix Renormalization Group (DMRG)
method.

For the sake of completeness we present a brief introduction to the
DMRG method. A superblock is called to a large block which contains
several smaller blocks, one of which is the block to be used in the
blocking procedure of the BRG method. Let us suppose that we have
diagonalized a superblock and thereby obtained one particular state
$|\psi \rangle$ which is called the {\em target state} and probably
will be the ground state. Let $\{ |i \rangle$, $i=1, \ldots, l \}$ be
a complete set of states of the the block B which we call  ``the
system". Let also  $\{ |j \rangle$, $j=1, \ldots, J$ be a complete
set of states of the superblock which we call  ``the universe" (see
Fig.5). Now we proceed to decompose the target state $|\psi \rangle $
into  its system- and universe-parts according to the following
equation,

\begin{equation} |\psi \rangle = \sum _{i,j} \psi _{ij} |i \rangle |j
\rangle       \label{d13} \end{equation}

\noindent Next we want to devise a procedure to produce a set of
states of the system  denoted by

\begin{equation} |u^{\alpha} \rangle, \ \ \alpha = 1,\ldots , n_B  \
\ \mbox{with} \ \
 |u^{\alpha} \rangle = \sum _i u^{\alpha}_i |i\rangle
\label{d14} \end{equation}

\noindent which are optimal for representing the target state $|\psi
\rangle$ in a sense to  be specified below. The number of states kept
is such that $n_B < l$ so that $|\psi \rangle$  is represented
approximately, that is,

\begin{equation} |\psi \rangle \approx |\tilde{\psi } \rangle :=
\sum _{\alpha,j} a_{\alpha,j} |u^{\alpha }\rangle \ |j \rangle
\label{d15} \end{equation}

\noindent where $a_{\alpha,j}$ are components to be determined by
demanding that the following  distance to be a minimun:

\begin{equation} {\cal D} := | |\psi \rangle - |\tilde{\psi } \rangle
|^2      \label{d16} \end{equation}

\noindent This minimization problem requires to vary over all
$a_{\alpha,j} $ and $u^{\alpha } $  subject to the condition,

\begin{equation} \langle u^{\alpha } | u^{\alpha' }\rangle  = \delta
_{\alpha \alpha'}   \label{d17} \end{equation}

\noindent White has proved that the solution to this minimization
problem is given by  the {\em optimal states} $u^{\alpha}$ being
eigenvectors of the reduced density matrix of the  system as part of
the universe whose eigenvalues are the $n_B$ largest in magnitude
\cite{white}.

\noindent The reduced density matrix
 for the system depends on the state of the universe which  in this
case is a pure state $| \psi \rangle $. Therefore, the density matrix
for the system is given by,

 \begin{equation} \rho _{i i'} = \sum _j \psi _{ij} \psi _{i'
j}                               \label{d27} \end{equation}

\noindent To summarize,  {\em when the entire lattice is assumed to
be in a pure state, the optimal states to be kept are then $n_B$ most
significant states of the reduced density matrix of the block,  say
B.}

Next step in putting density matrix RG ideas at work is to devise an
efficient algorithm  based upon these developments. A density matrix
algorithm is defined mainly by two features, the same as the standard
BRG algorithm is, namely, according to the form of the superblock and
the manner in which the  block are enlarged, e.g., doubling the block
B' = B B (Kadanoff) or adding a single site  B' = B + site (Wilson in
Kondo).

\noindent  As far as computer power is concerned, generally is more
efficient to enlarge the block by adding  a {\em single site} rather
than doubling the block. The reason for this is that the
diagonalization  of a superblock composed of say $p$ identical blocks
is difficult for a many-particle interacting  system for the
dimension of the Hilbert space of states goes like $n_B^p$, assuming
that  $n_B$ states are kept per block.

\noindent This has led White to propose a variety of density matrix
algorithms for both  finite and infinite size systems which rely on
Wilson's method for enlarging the system. Thus, they are
intrinsically one-dimensional methods.

\noindent In this paper we are presenting two new DMRG methods
(Variational and Fokker-Planck) which are based on the blocking
procedure of reducing the system size as  the standard Block
Renormalization Group method ot the previous section. This makes them
potentially well-suited to address higher dimensional systems.

 \noindent Moreover, the DMRG algorithms introduced by White are
intrinsically numerical for they are  based upon Wilson's procedure
of enlarging the system size. On the contrary, the BRG study  of the
ITF model carried out in  the previous section
 was done in an analytical fashion mainly because
 it is a blocking procedure. We shall present hereby a new treatment
of the  ITF model along the lines of the density matrix RG method
which is {\em analytical}.

\noindent To this purpose, we have incorporated the block method in
the DMRG algorithm. In addition, we shall incorporate another
ingredient for the algorithm to become analytical: we shall choose
the target state (ground state) using a variational method or a
Fokker-Planck  method, and then compare the results obtained in these
two fashions with the standard BRG  results of the previous section.

 The way to combine the above tools is as follows. We first set up a
variational ground state  for the {\em whole chain} whose energy is
determined by solving the corresponding minimization  equations. Next
we use this state to construct a block density matrix $\rho $ for a
block having  two sites in the philosophy of the BRG method. This
$\rho $ turns out to be (see below) a  $4 \times 4 $ matrix whose two
largests eigenvalues denoted by $|A \rangle$ and $|B \rangle$ are
kept to construct the intertwiner operator $T_0$ as in sections 2 and
3.
 Once we make contact with the  blocking method, the iteration
procedure goes over and over. The important point now is that we  can
keep the DMRG study at an analytical level.

It is worth noticing that we are introducing two new features in this
fashion. In the original DMRG  method of White the target state
selected comprises just a few sites of the chain while  now we are
using the whole chain. On the contrary, in doing so we have to resort
to variational  or Fokker-Planck methods to handle the problem, while
in the numerical DMRG the target  state selected is exact for the
particular size of the superblock chosen.

\subsection{Variational DMRG.}

In reference \cite{german-esteve} a new method based on the
combination of
 perturbative and variational techniques was presented to study
quantum lattice Hamiltonians. The general ideas of this method are
illustrated in the example of the Ising model in an  transverse field.
The method relies on the choice of an exponential ansatz  $\psi
(\lambda) \exp [U(\lambda)] \psi_0$, which is a sort of generalized
lattice version  of a Jastrow wave function. Perturbative and
variational techniques are used to get successive approximations of
the  operator $U(\lambda)$. Perturbation theory is used to set up a
variational method which in turn produces nonperturbative results.
This method allows one to associate to the original
quantum-mechanical problem a  statistical-mechanical system defined
in the same spatial dimension. In some instances these
statistical-mechanical systems turn out to be integrable, which
allows one to obtain exact upper bounds to the energy.
 We shall briefly review this method hereby but for a detailed
account we refer to \cite{german-esteve}.

   Let us suppose that we are given a hamiltonian of the form
$H(\lambda ) = H_0 + \lambda H_1$, where $H_0$ has a nondegenerate
ground state $\psi _0$ and $\lambda $ is a coupling constant. In
reference \cite{german-esteve} it was proposed to construct the
ground state  $\psi (\lambda )$ of $H(\lambda )$ as

\begin{equation} \psi (\lambda ) = \exp (\sum _{n=1}^{\infty} \lambda
^n U_n) \ \psi _0    \label{b1} \end{equation}

\noindent Solving perturbation theory in $\lambda $ to order $\nu $
implies  the knowledge of the collection of operators $\{ U_n
\}_{n=1, \ldots ,\nu }$.  Each operator $U_n$ consist in fact of a
sum of ``irreducible" operators $V_I$,

\begin{equation}
 U_n = \sum _I p_{n,I}
V_I
\label{b2} \end{equation}

\noindent Hence inserting (\ref{b2}) into (\ref{b1}) and
interchanging  the order of the sums one arrives to a ``dual"
description of the ground state

\begin{equation} \psi (\lambda ) = \exp (\sum _{I} \alpha _I (\lambda
) V_I) \ \psi _0    \label{b3} \end{equation}

\noindent where $\alpha _I (\lambda ) = \sum _n \lambda ^n p_{n,I}$.

  This expression suggests an alternative approximation to the
ground state $\psi (\lambda )$ which consists in choosing only a
class of irreducible operators $V_I$  whose weights $\alpha _I$ are
determined variationally. This was precisely the approach applied in
\cite{german-esteve} to the Ising model in a transverse field.

According to the perturbative-variational (PV) method
 \cite{german-esteve} we must first determine the form of the  set
operators $\{ U_n \}$ by inserting the exponential ansatz  (\ref{b1})
in the Schrodinger equation for $H$. Then these operators  serve us
to construct variational wave functions by inserting them  back in
the exponential ansatz. The perturbative equations that  determine
the $U_n$ for the lowest order in perturbation theory  obey the
equations,

\begin{equation} \mbox{order $\lambda$:} \ \ \ \ ([H_0,U_1] + H_1)
\psi _0 = E^{(1)} \psi _0         \label{b8b} \end{equation}

\begin{equation} \mbox{order $\lambda ^2$:} \ \ \ \ ([H_0,U_2] +
[H_1,U_1] + 1/2 ([[H_0,U_1],U_1])) \psi _0 = E^{(2)} \psi
_0                                         \label{b8c} \end{equation}

\noindent where $E^{(1)}$ and $E^{(2)}$ are the perturbative
energies to first and second order.

\noindent A solution of equations (\ref{b8b}) and (\ref{b8c}) in the
case where $H_0=-\sum_j \sigma_j^x$
 and $H_1=-\sum_j \sigma_j^z \sigma_{j+1}^z$  is the following,

\begin{equation} U_1 = \frac{1}{4}\sum_j \sigma_j^z
\sigma_{j+1}^z                                  \label{b9}
\end{equation}

\begin{equation} U_2 = \frac{1}{16} \{ \sum_j \sigma_j^z
\sigma_{j+2}^z    + \sum_j \sigma_j^z (\sigma_{j+1}^z +
\sigma_{j-1}^z) \} \label{b9b} \end{equation}

\noindent In the following we shall make use of the simplest
exponential ansatz based on  the $U_1$ operator only.

Now let us start searching for the block density matrix using the
variational method. Recall that the  density matrix in Eq.
(\ref{d27}) can be rewritten as a scalar product of two block states
$| \psi \rangle _B$ and  $| \psi' \rangle _{B'}$ given by their
defining system-universe decompositon  of the previous section:

\begin{equation}
 | \psi \rangle _B = \sum _{i\in B} \sum_{j\in B^c} \psi _{i j}
|i\rangle | j \rangle   \label{v1} \end{equation}

\begin{equation}
 | \psi' \rangle _{B'} = \sum _{i' \in B'} \sum_{j' \in \mbox{B'}^c}
 \psi _{i' j'} |i' \rangle | j'
\rangle
\label{v2} \end{equation}

\noindent where $i$, $i'$ are system-indexes, $j$, $j'$ are
universe-indexes, $B$, $B'$ are  system-blocks and $B^c$,
$\mbox{$B'$}^c$ are universe-blocks. The density matrix $\rho$ is
given by the scalar product of these two block states:

 \begin{equation} \rho   = _B\hspace*{-1pt} \langle \psi | \psi'
\rangle_{B'}                \label{v3} \end{equation}

\noindent for upon substitution of Eqs. (\ref{v1}), (\ref{v2}) we
arrive at

 \[
 \rho   = \sum_{ij} \sum_{i'j'} \langle i| \langle j| \psi_{ij}
\psi_{i'j'} | i' \rangle | j' \rangle      \] \begin{equation}
 = \sum_{i' i}  | i' \rangle \langle i|    \rho_{i
i'}                                \label{v4} \end{equation}

\noindent with $\rho_{i i'}  $ as in Eq.(\ref{d27}).

Now let us consider  the following variational state ansatz for the
ground state  $| \psi _0 \rangle_{1,2,3,\ldots,N}$ of the ITF model
in a lattice with $N$ sites:

 \begin{equation} | \psi _0 \rangle_{1,2,3,\ldots,N} = \exp
(\frac{\alpha}{2}  \sum_{j=1}^N \sigma_j^z \sigma_{j+1}^z) \ | 0
\rangle_{1,2,3,\ldots,N}              \label{v5} \end{equation}

\noindent where $\alpha $ is a variational parameter which is
determined by minimizing the  vacuum expectation value of the ITF
Hamiltonian in this state, thereby making the parameter to become a
function  $\alpha = \alpha (J/2\Gamma)$ of the coupling constant of
the model; and  $| 0 \rangle_{1,2,3,\ldots,N} = | 0 \rangle_1 \otimes
\ldots \otimes | 0 \rangle_N$  with $| 0 \rangle  $ being the ground
state of the $\sigma^x $ matrix. This exponential-variational  ansatz
constitutes part of a method called Perturbative-Variational method
(PV) developed in  \cite{german-esteve} for spin systems and in
\cite{pv-germanyo} for fermionic systems. The PV method is
essentially a cluster method which combines perturbative and
variational techniques. Using Eq. (\ref{v3}) we construct the block
density matrix out of this target state for a block  containing 2
sites as:

 \begin{equation} \rho^{PV} = \frac{1}{Z_N(\alpha )} \ _{1,2}
\hspace*{-1pt}  \langle \psi_0 | \psi_0
\rangle_{1',2'}                                         \label{v6}
\end{equation}

\noindent where $Z_N(\alpha )$ is a normalization factor being the
norm of the state which can  be interpreted as the partition function
of a certain associated statistical model  \cite{german-esteve}. As a
matter of fact, in the large $N$ limit, it turns out to be simply:

 \begin{equation} Z_N(\alpha ) = \langle \psi_0 | \psi_0 \rangle =
(\cosh \alpha )^N, \ \ N  \gg 1    \label{v7} \end{equation}

\noindent Inserting the ansatz (\ref{v5}) into (\ref{v6}) we can
express the density matrix in a more  appealing form, namely,

  \begin{equation} \rho^{PV} = \frac{1}{Z_N(\alpha )} \ _{1,2}
\hspace*{-1pt}  \langle 0 | \exp \frac{\alpha }{2}
(\mbox{$\sigma$}_1^z \mbox{$\sigma$}_2^z +  \mbox{$\sigma'$}_1^z
\mbox{$\sigma'$}_2^z) \ Z_{N-2}^{(0)} (\alpha, h, h')  | 0
\rangle_{1',
2'}
\label{v8} \end{equation}

\noindent where we have defined the following quantities,

  \begin{equation}
 Z_{N-2}^{(0)} (\alpha, h, h') := _{3,\ldots,N} \hspace*{-1pt}
\langle 0 |
 \exp^{(\alpha \sum_{j=3}^{N-3} \sigma_j^z \sigma_{j+1}^z +
 h \sigma_3^z + h'  \sigma_N^z)} | 0 \rangle_{3,\ldots,N}
\label{v9} \end{equation}

  \begin{equation} h := \frac{\alpha}{2} (\sigma_2^z +
\mbox{$\sigma'$}_2^z)   \label{v10} \end{equation}

  \begin{equation} h' := \frac{\alpha}{2} (\sigma_1^z +
\mbox{$\sigma'$}_1^z)   \label{v11} \end{equation}

\noindent For the time being, it is convenient to shift $N-2
\rightarrow N$ in order to make  expressions easier (at the end we
shall come back to the correct value).

\noindent It is possible to recast $Z_{N-2}^{(0)} (\alpha, h, h') $
into the following form,

\[
 Z_{N}^{(0)} (\alpha, h, h') = _{1,\ldots,N} \hspace*{-1pt} \langle 0
|
 \exp (\alpha \sum_{j=1}^{N-1} \sigma_j^z \sigma_{j+1}^z +
 h \sigma_1^z + h'  \sigma_N^z) | 0 \rangle_{1,\ldots,N}      \]
  \begin{equation}
 = \frac{1}{2^N} \sum_{\{ \sigma_1,\ldots, \sigma_N \} }  \
   \exp (\alpha \sum_{j=1}^{N-1} \sigma_j \sigma_{j+1} +
 h \sigma_1 + h'
\sigma_N)                                               \label{v12}
\end{equation}

\noindent Now we can recognize this equation as the partition
function for the Ising model on an  open chain of $N$ sites subject
to an external magnetic field applied only at the ends of the chain.
 This partition function can be worked out exactly using standard
transfer matrix calculations  yielding the result:

\[ Z_N^{(0)} = \frac{1}{2^N} \{ [\cosh (h + h') + \cosh (h - h')]  (2
\cosh \alpha)^{N-1}  \]
  \begin{equation} + [\cosh (h + h') - \cosh (h - h')]  (2 \sinh
\alpha)^{N-1}  \}    \label{v13} \end{equation}

\noindent In the $N \rightarrow \infty$ limit in which we are
interested in, it further simplifies to:

  \begin{equation} Z_{N\rightarrow \infty}^{(0)} =  \frac{1}{2}
(\cosh \alpha)^{N-1}    [\cosh (h + h') + \cosh (h - h')]
\label{v14} \end{equation}

\noindent Inserting now Eq.(\ref{v14}) in Eq.(\ref{v8}) we arrive at
the following expression for the  PV block density matrix:

\[ \rho ^{PV} = \frac{1}{2 (\cosh \alpha)^3}  _{1, 2} \hspace*{-1pt}
\langle 0 | \exp \frac{\alpha}{2}  (\sigma_1^z \sigma_2 +
\mbox{$\sigma'$}_1^z \mbox{$\sigma'$}_2^z) \]
   \begin{equation} \times  [\cosh \frac{\alpha}{2} (\sigma_1^z +
\sigma_2^z +  \mbox{$\sigma'$}_1^z  + \mbox{$\sigma'$}_2^z) +
 \cosh \frac{\alpha}{2} (\sigma_1^z - \sigma_2^z +
\mbox{$\sigma'$}_1^z  - \mbox{$\sigma'$}_2^z)]  | 0 \rangle_{1',
2'}                  \label{v15} \end{equation}

\noindent This is a nice result. Observe that the piece  $_{1, 2}
\hspace*{-1pt} \langle 0 | \exp ^{\frac{\alpha}{2}  (\sigma_1^z
\sigma_2)} \  \exp^{(\mbox{$\sigma'$}_1^z \mbox{$\sigma'$}_2^z)}  | 0
\rangle_{1', 2'}   $ corresponds to a  density matrix of a pure state
$\rho = | \phi \rangle_{1', 2'} \ _{1,2} \hspace*{-1pt} \langle \phi
|$ is  the projection of the target state $\psi _0$ onto the block
$(1,2)$. The extra terms in  Eq.(\ref{v15}) are the novel features
that the DMRG(PV) method brings about.

\noindent To proceed further and give $\rho^{PV}$ a simple matricial
form, it is convenient to change  basis from eigenstates $| 0
\rangle$, $| 1 \rangle$ of $\sigma^x$ to eigenstates  $| + \rangle$,
$| - \rangle$ of $\sigma^z$. The notation is,

   \begin{equation}
 | + \rangle = \left( \begin{array}{c} 1 \\  0  \end{array} \right) ,
\ \   | - \rangle = \left( \begin{array}{c} 0 \\  1
\end{array}      \right)
\label{v16} \end{equation}

   \begin{equation}
 | 0 \rangle = \frac{1}{\sqrt{2}} \left( \begin{array}{c} 1 \\  1
\end{array} \right) , \ \   | 1 \rangle =  \frac{1}{\sqrt{2}} \left(
\begin{array}{c} 1 \\  -1 \end{array}
\right)                                        \label{v17}
\end{equation}

\noindent In the new basis $\{ | + \rangle, | - \rangle \}$ the
components of $\rho ^{PV}$ are:

\[ \rho^{PV}_{\mbox{$\sigma'$}_1 \mbox{$\sigma'$}_2  \mbox{$\sigma
$}_1 \mbox{$\sigma $}_2} =  \frac{1}{8 (\cosh \alpha)^3}
\exp{\frac{\alpha }{2} (\sigma_1 \sigma_2 +  \mbox{$\sigma'$}_1
\mbox{$\sigma'$}_2)} \]
   \begin{equation} \times  [\cosh \frac{\alpha}{2} (\sigma_1 +
\sigma_2 +  \mbox{$\sigma'$}_1  + \mbox{$\sigma'$}_2) +
 \cosh \frac{\alpha}{2} (\sigma_1 - \sigma_2 +  \mbox{$\sigma'$}_1  -
\mbox{$\sigma'$}_2)]                  \label{v18} \end{equation}

\noindent Now $\rho^{PV}$ takes the following matricial form in the
basis  $\{ + +, - -, + -, - + \}$:

   \begin{equation} \rho^{PV} = \frac{1}{4 (\cosh \alpha)^3} \left(
\begin{array}{cccc} e^{\alpha } (\cosh \alpha)^2 & e^{\alpha } &
\cosh \alpha & \cosh \alpha \\ e^{\alpha } & e^{\alpha } (\cosh
\alpha)^2 & \cosh \alpha & \cosh \alpha \\
 \cosh \alpha & \cosh \alpha & e^{-\alpha } (\cosh \alpha)^2 &
e^{-\alpha } \\ \cosh \alpha & \cosh \alpha & e^{-\alpha } &
e^{-\alpha } (\cosh \alpha)^2 \end{array}
\right)
\label{v19} \end{equation}

\noindent We can readly check that $\rho^{PV}$ is normalized:

   \begin{equation} tr \rho^{PV} =
1                                               \label{v20}
\end{equation}

\noindent Next step in the DMRG algorithm is to diagonalize the
density matrix (\ref{v19}) and  truncate to the largest ones; in this
case the truncation is to two states to be denoted by  $| A \rangle $
and $| B \rangle $ and will play the parallel role of the states   $|
G \rangle $ and $| E \rangle $ for the block Hamiltonian $H_B$ of
Section 3.

\noindent We do not need to make a ``blind" diagonalization of this
$4\times4$ matrix for we may  take advantage of what we have learnt
in Section 3 about the eigenstates of the $H_B$ in the  ITF model.
Then, the largest eigenvector say $| A \rangle$ will be in the even
sector  $\{ | 0 0 \rangle, | 1 1 \rangle \}$, while the next to the
largest one, say $| B \rangle $, will be in the  odd sector $\{ | 0 1
\rangle, | 1 0 \rangle \}$. According to this analysis, we may write
those states  as,

   \begin{equation} | A \rangle =  x_{00} | 0 0 \rangle + x_{11}  | 1
1 \rangle    \label{v21} \end{equation}

   \begin{equation} | B \rangle =  x_{01} | 0 1 \rangle + x_{10}  | 1
0 \rangle    \label{v22} \end{equation}

\noindent where $x_{00}, x_{11}, x_{01}, x_{10} $ are the components
to be determined. Expressing the states $| 0 0 \rangle$, $ | 1 1
\rangle $  $\ldots $ in the basis of $\{ |+\rangle, |-\rangle \}$,
the diagonalization of the density matrix (\ref{v19}) yields the
following eigenvalues:

   \begin{equation} w_0 = \frac{1}{4 (\cosh \alpha)^3} [\cosh \alpha
(1 + \cosh ^2\alpha ) +  \sqrt{\cosh^2 \alpha (1 + \cosh ^2\alpha )^2
- \sinh^4 \alpha }]                \label{v23} \end{equation}

    \begin{equation} w_1 = \frac{1}{4 (\cosh \alpha)^3} [\cosh \alpha
(1 + \cosh ^2\alpha ) -  \sqrt{\cosh^2 \alpha (1 + \cosh ^2\alpha )^2
- \sinh^4 \alpha }]                \label{v24} \end{equation}

    \begin{equation} w_3 = \frac{1}{4 (\cosh \alpha)^3} e^{\alpha}
\sinh^2 \alpha             \label{v25} \end{equation}

    \begin{equation} w_4 = \frac{1}{4 (\cosh \alpha)^3} e^{-\alpha}
\sinh^2 \alpha             \label{v26} \end{equation}

\noindent For $\alpha $ small it is easy to see that the eigenvalues
are sorted according to

    \begin{equation} w_0 > w_3 > w_1 > w_4              \label{v27}
\end{equation}

\noindent For arbitrary values of $\alpha $, which in turn amounts to
arbitrary values of the  coupling constant $g=J/2\Gamma $ due to the
variational equations to be given bellow, we have plotted these 4
eigenvalues in Fig. 6 observing that there are not level crossings in
the  whole range of variation of $\alpha $ and that the sorting in
Eq.(\ref{v27}) holds all over, not just  for small $\alpha$. It is
very important for our DMRG(PV) method to work properly that this
property holds up, for when we truncate to the eigenstates of the
largest eigenvalues, $w_0$ and  $w_3$, the physics will not change
qualitatively when varying the coupling constant $g$.

\noindent Moreover, it is also possible to show that the eigenvectors
$|A \rangle$ (of $w_1$)  and $|B \rangle$ (of $w_3$) are given by:

 \begin{equation} |A \rangle = \frac{|0 0 \rangle + a^{PV} |1 1
\rangle}{\sqrt{1 + (a^{PV})^2}}    \label{v28} \end{equation}

 \begin{equation} |B \rangle = \frac{|0 1 \rangle +  |1 0
\rangle}{\sqrt{2}}    \label{v29} \end{equation}

\noindent with the function $a^{PV}$ now the following function of
the parameter $\alpha $ (after some tedious algebra),

 \begin{equation} a^{PV} = \frac{\sqrt{\cosh^2 \alpha (1 + \cosh^2
\alpha)^2  - \sinh^4 \alpha} - 2\cosh \alpha} {\sinh \alpha (1 +
\cosh^2 \alpha
)}
\label{v30} \end{equation}

\noindent Notice that these states which now come from a DMRG(PV)
analysis have the same  form as the states $|G \rangle$, $|E \rangle
$ of the block Hamiltonian $H_B$  Eqs.(\ref{38a})-(\ref{38b}), the
difference being  in the dependence of the function $a$ upon the
coupling constant. This means that we have again  the same structure
as in the BRG analysis where the intertwiner operator $T_0$ was fully
determined, and consequently the whole RG procedure, by a single
function $a=a(g)$ of the coupling  constant. To obtain $a^{PV} =
a^{PV}(g)$ we need the variational equation relating $g$ with
$\alpha $. This can be found in \cite{german-esteve} and is given by,

 \begin{equation} g = \frac{J}{2 \Gamma} =  \tanh
\alpha                                          \label{v31}
\end{equation}

\noindent Observe that when $J/ \Gamma \ll 1$ then $\alpha =
J/2\Gamma $ and thus $a^{PV} \sim \alpha/2 = J/4\Gamma$. Now it is
possible to eliminate the intermediate parameter $\alpha $ between
Eqs. (\ref{v30}) and (\ref{v31}) yielding the desired formula,

  \begin{equation} a^{PV}(g) = \frac{\sqrt{1 - g^2 + \frac{g^6}{4}} -
(1 - g^2)}{g (1 - \frac{g^2}{2})}   \label{v32} \end{equation}

\noindent Therefore we may appreciate that this function shares the
same qualitatives properties as $a_{BRG}$ (\ref{40})
 does, Eq.(\ref{49}) ,namely it goes to zero linearly when $g
\rightarrow 0$ and it is bounded  below 1, i.e., $0 < a^{PV} < 1$ for
$0 < g < 1$. It addition, it has a singularity at $g = \sqrt{2}$
which prevents the extension of this method beyond that singular
point. The origin of this singularity  is due to the variational
nature of the method (see \cite{german-esteve}). Nevertheless the
critical  region lies within the region of applicability of the PV
method.

\noindent Furthermore, it is also possible to define a variational
DMRG method valid for the whole  range of variation of the coupling
constant. To do this, we simply recall that for small $\alpha$ the
coupling constant depends linearly on the variational parameter,

  \begin{equation}
 g \sim \alpha , \ \  \  \ \mbox{for $\alpha$ small}   \label{v33}
\end{equation}

\noindent Thus, we may define another function say $a^{PV'}(g)$ by
simply substituting $\alpha$ by $g$ in $a(\alpha)$ (\ref{v30}),

 \begin{equation} a^{PV'} = \frac{\sqrt{\cosh^2 g (1 + \cosh^2 g)^2
- \sinh^4 g} - 2\cosh g} {\sinh g (1 + \cosh^2 g
)}
\label{v34} \end{equation}

\noindent In this fashion, $a^{PV'}(g)$ shares the same properties
with $a_{BRG}$, without  any singularity in the range of $g$. In
fact, $a^{PV'}(g)$ has a horizontal asimptota at 1 as  $a_{BRG}$ does
(see Fig  7).

\subsection{Fokker-Planck DMRG.}

This is another approximated version for preparing the  target state
to be projected onto the block-system in order to construct another
analytical DMRG  method based upon a blocking procedure. The details
of how the Fokker-Planck (FP) method  is applied to construct an
approximate version of the real ground state of the ITF model are
given in the  original paper  \cite{german-fernando}. Briefly stated,
what the FP method does is to start with an exponential ansatz as in
the variational method Eq. (\ref{v5}), but instead the parameter
$\alpha $ is fixed by demanding that a certain  Fokker-Planck
Hamiltonian $H^{FP}$ (to be determined along the way) satisfies the
eigenvalue  equation to a certain order $\nu$ in the perturbative
expansion of the parameter $\alpha $, that is,

 \begin{equation} H^{FP}(\alpha) \Psi^{FP}(\alpha ) = E^{FP}(\alpha )
\Psi^{FP}(\alpha )  \label{v35} \end{equation}

\noindent such that the exact ITF Hamiltonian $H_{ITF}$ and its
Fokker-Planck approximation version $H^{FP}(\alpha) $ differ in
operators say $V_I$ which involve interactions between lattice  sites
a $\nu +1 $ distant apart or larger. Squematically,

 \begin{equation} H_{ITF} - H^{FP}(\alpha) = \sum_{I> \nu} C_I(\alpha
) V_I   \label{v36} \end{equation}

\noindent Correspondingly,  the FP-energy $ E^{FP}(\alpha )$,
although incorporating non-perturbative effects, should agree with
the exact ground state energy up to order $\nu +1$ in $\alpha$. In a
sense, this gives the best ``exact" aproximation to the Hamiltonian
$H_{ITF}$ to order $\nu $ in perturbation  theory
\cite{german-fernando}.

\noindent It is possible to show that these conditions fix the
relationship between $\alpha $ and $g$. To lowest order this is given
by \cite{german-fernando},

 \begin{equation} g = \frac{1}{2} \sinh (2 \alpha
)                                                  \label{v37}
\end{equation}

\noindent Again we see that for small coupling constant $\alpha$ and
$g$ are equal as in the  variational method.

In order to obtain the Fokker-Planck function, say $a^{FP}(g)$, we
first notice that as in this FP  method we start with the same
exponential ansatz (\ref{v5}) as in the variational method, the same
function $a(\alpha)$ in Eq. (\ref{v30}) is valid here. The new
feature is that we have to use the  relation (\ref{v37}) now to
express the function $a^{FP}(g)$ in terms of the coupling constant.
After some tedious algebra we arrive at the following expression:

 \begin{equation} a^{FP}(g) = \frac{\sqrt{(8 + 26 g^2 + 4 g^4) + (8 +
6g^2) \sqrt{1 + 4g^2}} - 2(1 + \sqrt{1 + 4 g^2})} {g (3 + \sqrt{1 +
4g^2})}
\label{v38} \end{equation}

\noindent This function contains all the information which upon
inserted in the intertwiner operator  $T_0$ gives rise to what we
denote by a DMRG(FP) method. By looking at Fig.7 we notice again
that  $a^{FP}(g)$ has the same qualitative properties as $a_{BRG}$
does.

In this section we have introduced 3 functions $a(g)$ namely,
$a^{PV}(g)$, $a^{PV'}(g)$ and  $a^{FP}(g)$, related to different
analytical realizations of the density matrix RG ideas. It is our
purpose now to check the goodness of those methods by comparing their
predictions for the  critical exponents with the exact values
already found by the standard BRG method  in Sect.3.

In Fig.7 we have plotted the 4 functions $a(g)$ along with the
universal function $a(g_c)$ introduced  in Eq. (\ref{51})
 whose cuts with the functions $a(g)$ gives the predictions on the
location of the critical  point $g_c$ for each method. Recalling that
the exact value is $g_c^{exact} = 1/2$, we see that the  closest
value to this one is produced by the Fokker-Planck version of the
DMRG method,  even better than the standard BRG. Nevertheless, we may
appreciate from Fig.7 that the 4 methods  lie rather close to one
another within the critical region, the major differences being
present off  criticality when entering the strong coupling region.
The particular values of $g_c$ are gathered in  Table 2.

Another interesting function to be plotted  is the beta function
$\beta (g)$ obtained for each method according to the analysis of
Sect.3, Eq. (\ref{53b}). We show these results in Fig.8  where we
observe that the 3 new beta functions introduced in this  section by
means of variational and Fokker-Planck DMRG methods have the same
qualitative  behaviour as the standard BRG beta function of Sec.3. We
know that in particular this means that  the unstable character of
the fixed point $g_c$ is preserved by these new methods. Moreover,
we  notice again that in the critical region the differences are
small, namely, the cut with the $g$-axis  and the slope at the
corresponding $g_c$. These two latter properties are related to
critical exponents.

As far as the critical exponents is concerned, we have collected in
Table 2 the exponents computed  previously for the standard BRG
method: correlation length exponent $\nu$, dynamical exponent $z$,
magnetic exponent $\beta $ and the gap exponent $s$. Notice first
that the 3 new DMRG  methods also satisfy the scaling relation,

\[ \beta = \frac{1}{2} z \nu \]

\noindent which we know does not hold for the exact  solution of the
ITF model.

\noindent From Table 2 we arrive at the following conclusions,

\begin{itemize}

\item The best correlation length exponent $\nu $ is provided by the
DMRG Perturbative-Variational method.

\item The best dynamical exponent $z $ is provided by the DMRG
Fokker-Planck method.

\item The best magnetic exponent $\beta $ is provided by the DMRG
Perturbative-Variational method. \item The best gap exponent $s $ is
provided by the DMRG Fokker-Planck method.

\end{itemize}

\noindent From these results we may draw the conclusion that the RG
methods based on  block density matrix, either variational or
Fokker-Planck, provide an improvement respect to  the standard BRG
methods, though it is not a major improvement. One of the reasons
why  this improvement is not as good as the numerical results
obtained by White \cite{white} relies on the  fact that we have just
kept 2 states in our analytical DMRG method while in the numerical
treatment quite a lot states are kept.

\section{Conclusions}

We have presented in this paper an analytic formulation of the
recently proposed Density Matrix RG method \cite{white}. This method
was originally developed in a numerical fashion mainly because it
relies on the Wilsonian procedure of enlarging lattice sizes in the
real-space RG. As this Density Matrix RG method has become a powerful
tool to  compute static and dynamic  properties of quantum lattice
systems at zero  temperature, we find interesting to devise an
analytic formulation of this method to be tested against the old
procedures based upon the standard Block RG method introduced by
Drell et al. \cite{drell} to study QCD.

The new feature of the DMRG method is its ability to take into
account the unavoidable interaction between the block selected  for
truncating states and the rest of the lattice. This feature is the
more relevant the more strongly correlated is the system under study,
such as lattice QCD and strongly  correlated electrons in High-$T_c$
materials.

We have been able to devise an analytic formulation of the DMRG
method by combining the idea of ``interacting blocks" (that
is,``non-isolated") with two other approaches. One approach  is the
old idea of blocking ``a la Kadanoff" instead of the  Wilsonian
procedure. To this end we have presented an unified  formulation of
the old BRG method (sect.3) in terms of what we call the intertwiner
operator. This facilitates the task of bringing together the Kadanoff
blocking with the DMRG. The second approach has been to devise an
analytical method to produce ``target states" which are the basic
ingredients in the DMRG calculations. To this purpose we have used
two  recently proposed methods to deal with quantum lattice
Hamiltonians: the Perturbative-Variational method
\cite{german-esteve} and  the Fokker-Planck method
\cite{german-fernando}. We have coined the names DMRG(PV) and
DMRG(FP), respectively, for the density matrix RG methods coming out
of these two approaches \footnote{After we finished this work we have
been aware of  references \cite{jap} where those authors also propose
new extensions of the DMRG method.}.

In order to facilitate the task of testing these two new DMRG
methods, we have chosen the simple ITF model where previous  work
with the BRG method is available. The results of computing several
critical exponents are described in Sect.4 where we have seen that
 either  of the new DMRG methods perform better than the old BRG,
although the improvement we get is not as good as White's numerical
DMRG for we keep a much lower number of states during the truncation
among other reasons.

In addition, there is another line of work in order to improve  our
analytical DMRG methods. This is  via the exact diagonalization of
the block density matrix corresponding  to 4 lattice sites (starting
from the exact ground state for 4 lattice sites) and then to proceed
in the  usual fashion to construct the corresponding new $a(g)$
function which would allow us to  carry the blocking method. We leave
this posibility  open for future work.

Another reason to seek analytical formulations of the DMRG method is
the possibility of generalizing the current one-dimensional
algorithms to higher dimensions. We consider our DMRG(PV) and
DMRG(FP) methods as a first attempt at this goal. As a matter of
fact, it is possible to use them to prepare 2-dimensional target
states in the ITF model and  to proceed with a DMRG analysis.
Moreover, as far as  fermion systems is concerned, it is also
possible to apply the  new density matrix methods using the
perturbative-variational  techniques described in \cite{pv-germanyo}.

Finally, the renormalization group method is one of the basic
concepts in several branches  of Physics and we believe that  we are
currently facing a reconsideration of the old renormalization group
ideas which will certainly have implications in areas such as Field
Theory, Statistical Mechanics and Condensed Matter.


\vspace{20 pt}

{\bf Acknowledgements}
 \vspace{20 pt}

Work partially supported in part by CICYT under  contracts AEN93-0776
(M.A.M.-D.) and PB92-1092, European Community Grant ERBCHRXCT920069
(G.S.).

It is a pleasure to thank A. G\'onzalez-L\'opez for sharing with  us
his access to the Alpha computer {\em Ciruelo} to make  some of the
numerical computations in this work.



\def\baselinestretch{1.5} \noindent  %

\newpage \section*{Table captions}

{\bf Table 1 :} Block Renormalization Group Method BRG1 ($n_B$,$k'$).

{\bf Table 2 :} Critical exponents for the ITF model according to
different RG methods and exact solution.

\newpage \section*{Figure captions} \noindent

 {\bf Figure 1:} Pictorical decomposition of the Hamiltonian $H$ into
single-site part  $H_S$ and two-nearest-neighbour-site part $H_{SS}$.

 {\bf Figure 2 :}  Block decomposition of the open chain into blocks
with $n_B=3$ sites.

 {\bf Figure 3 :} Pictorical representation of the block Hamiltonian
$H_B$ and the  interblock Hamiltonian $H_{BB}$ for the ITF model.

 {\bf Figure 4 :}  Pictorical representation of the truncation
procedure for the block and  interblock Hamiltonians in the ITF model.

{\bf Figure 5  :} Lattice decomposition into ``system"- and
``universe"-parts.

{\bf Figure 6 :} The 4 eigenvalues $w_0$, $w_1$, $w_3$ and $w_4$ of
the block density matrix corresponding to the  variational DMRG
method.

 {\bf Figure 7 :} Plot of the functions $a(g)$ according to the
methods: BRG (solid line), DMRG(PV) (grey line), DMRG(FP) (dashed
line), DMRG(PV').

{\bf Figure 8 :} Plot of the beta functions $\beta (g)$ according to
the methods: BRG (solid line), DMRG(PV) (grey line), DMRG(FP) (dashed
line), DMRG(PV').

\newpage

\begin{table} \caption{Block Renormalization Group Method BRG1
($n_B$,$k'$)} \vspace{2pt} \begin{tabular}{lc} \hline
\multicolumn{1}{l}{\rule{0pt}{12pt}
                Steps of the BRG1 Method}\\[2pt]
\hline\rule{0pt}{12pt} 1) Blocking Transformation: $H = H_B + H_{BB}$
& \\ 2) Diagonalization of $H_B$& \\ 3) Truncation within each block:
$T_0$& \\ 4) Renormalization of $H_B$ and $H_{BB}$& \\ \ \ \ $
H'_{s'} = T^{\dag}_0 H_B T_0$& \\ \ \ \ $ H'_{s's'} = T^{\dag}_0
H_{BB} T_0$& \\ 5) Iteration: Repeat $1) \rightarrow 4)$ for $H' =
H'_{s'} + H'_{s's'}$& \\[2pt] \hline \end{tabular} \end{table}

\begin{table} \caption{Critical exponents for the ITF model according
to different RG methods and exact solution.} \vspace{2pt}
\begin{tabular}{ccccccl} \hline \multicolumn{1}{c}{\rule{0pt}{12pt}
       Method}&\multicolumn{1}{c}{
       $g_c$}&\multicolumn{1}{c}{
       $\nu$}&\multicolumn{1}{c}{
       $z$}&\multicolumn{1}{c}{
       $\beta$}&\multicolumn{2}{c}{
       $s$}\\[2pt] \hline\rule{0pt}{12pt} BRG & $0.3916$ & $1.4820$
& $0.5515$  & $0.4086$ & $0.8173$ & \\ DMRG(PV) & $0.3790$ & $1.4073$
& $0.5353$ &  $0.3767$ & $0.7534$ & \\ DMRG(FP) & $0.4011$ &
$1.5177$  & $0.5647$  &  $0.4285$ & $0.8570$ & \\ DMRG(PV') &
$0.3874$ & $1.4579$  & $0.5459$  & $0.3980$ & $0.7958$ & \\ Exact
Solution & $0.5$ & $1$ & $1$& $0.125$ & $1$ & \\[2pt] \hline
\end{tabular} \end{table}

\end{document}